\pdfoutput=1
\documentclass[12pt,fleqn]{article}

\usepackage{amsfonts,ulem,url} \usepackage{amssymb} \usepackage{amsmath}
\usepackage{amstext,verbatim,graphicx,ifthen,multirow,amsthm, subcaption, float, placeins}

\usepackage{natbib}
\usepackage[hidelinks]{hyperref}

\renewcommand{\thepage}{}
\renewcommand{\appendix}{\footnotesize\parindent 0cm\setcounter{equation}{0} 
\renewcommand{\theequation}{A.\arabic{equation}}
\setcounter{lemma}{0}\renewcommand{\thelemma}{A.\arabic{lemma}}}

\def\monthname{\ifcase\month\or
January\or February\or March\or April\or May\or June\or
July\or August\or September\or October\or November\or December\fi}

\renewcommand{\appendix}{\small\parindent 0cm\setcounter{equation}{0} 
\renewcommand{\theequation}{A.\arabic{equation}}
\setcounter{lemma}{0}\renewcommand{\thelemma}{A.\arabic{lemma}}
\setcounter{theorem}{0}\renewcommand{\thetheorem}{A.\arabic{theorem}}}
\graphicspath{{./}}

\usepackage{fancyhdr}
\def\secondpage{\clearpage\null\vfill
\pagestyle{empty}
\begin{minipage}[b]{0.9\textwidth}
\footnotesize\raggedright
\setlength{\parskip}{0.5\baselineskip}
Copyright \copyright \the\year\ comScore \par
This work is licensed under Creative Commons Attribution license (CC BY 4.0) license. 
\end{minipage}
\vspace*{2\baselineskip}
\cleardoublepage
\rfoot{\thepage}}

\makeatletter
\g@addto@macro{\maketitle}{\secondpage}
\makeatother

\begin{document}
\title{Generating Random Samples from Non-Identical Truncated Order Statistics\thanks{%
{\small We are grateful for comments by Hattie Young. This research was generously supported by Comscore.
%We are grateful for comments by  participants in the econometrics lunch seminar at Harvard 
%University, and discussions with Gary Chamberlain. 
%We are grateful to XXX for making the data available.
}}}
\author{Tyler Morrison\thanks{{\small tmorrison@comscore.com }} \and
 Sean Pinkney\thanks{{\small spinkney@comscore.com.}}
}
\date{
 \ifcase\month\or
January\or February\or March\or April\or May\or June\or
July\or August\or September\or October\or November\or December\fi \ \number%
\year
}
\maketitle

\begin{abstract}
We provide an efficient algorithm to generate random samples from the bounded $k$-th order statistic in a sample of independent but not necessarily identically distributed random variables. The bounds can be upper or lower bounds and need only hold on the $k$-th order statistic. Furthermore, we require access to the inverse CDF for each statistic in the ordered sample. The algorithm is slightly slower than rejection sampling when the density of the bounded statistic is large, however, it is significantly faster when the bounded density becomes sparse. We provide a practical example and a simulation that shows the superiority of this method for sparse regions arising from tight boundary conditions and/or over regions of low probability density. 

%SA added sentence
\end{abstract}

%\textbf{JEL Classification: C14, C21, C52}

%\textbf{Keywords:\  Potential Outcomes}

\baselineskip=20pt\newpage \setcounter{page}{1}\renewcommand{\thepage}{[%
\arabic{page}]}\renewcommand{\theequation}{\arabic{section}.%
\arabic{equation}}

\section{Introduction}
We consider the problem of how to efficiently draw pseudo-random variates from a truncated distribution where the distribution is equivalent to the $k$-th order statistic of a set of $N$ independent random variables. The CDF and inverse CDF for each of the $N$ distributions is known. The main contribution of this paper is an algorithm that bounds the $N$-dimensional hypercube so that the inverse CDF sampling method may be used (see \cite{devroye2006nonuniform}). 

Consider $N$ independent random variables $X_i$ whose CDFs, $F_{X_i}$, and inverse CDFs, $F_{X_i}^{-1}$, are known, and we want to draw a random sample from their $k$-th order statistic, $Y = X_{(k)}$ satisfying bounds of $A < Y < B$. As the bounds narrow, rejection sampling struggles to find acceptable samples. Instead of waiting for acceptable samples, we map the samples from the $N$-dimensional unit hypercube onto a restricted subset of another $N$-dimensional unit hypercube that is normalized to represent the restricted probability space. We then apply the inverse CDF to obtain our draws from $X_i$.

The procedure is useful in a range of applications that support the factorization of a distribution function into $N$ known distribution functions not necessarily identical. A frequent example is drawing bounded variates from the minimum or maximum of a sample of order statistics, which arise when modeling censored or truncated data. The 2-dimensional minimum or maximum case is presented as a practical motivating example prior to the general bounded $k$-th order statistic. 

\section{Two-Dimensional Case}

Define $Y = \min(X_1, X_2)$ where each $X_i$ are independent, and $Y$ is bounded below by $A$ and above by $B$. Further, assume we have access to their CDFs, $F_{X_i}$, and inverse CDFs, $F_{X_i}^{-1}$. Without the bound restriction, the sample is easily generated by drawing two independent uniform draws from $[0, 1]$, applying the inverse CDF to each draw and returning the minimum of the two. To satisfy the boundary condition, we restrict the region as 
$$
C^{\prime} = ([a_1, 1]\times [a_2, 1]) \setminus ([b_1, 1]\times [b_2, 1])
$$
where $a_i = F_{X_i}(A)$ and $b_i = F_{X_i}(B)$. Since $Y$ is defined as a minimum, the lower bound does not provide an issue; we simply require that each of the $X_i$ be greater than $A$. The upper bound requires constraining the space such that it is possible for one of the $X_i$'s to exceed $B$ as long as the other does not. As mentioned above, rejection sampling may be used, but bounds resulting in low-density regions suffer from long running times. Our method draws samples from the unrestricted unit square and maps the results onto the restricted subspace.

\begin{figure}[h]
	\includegraphics[trim={3cm 18cm 2.5cm 2cm},clip]{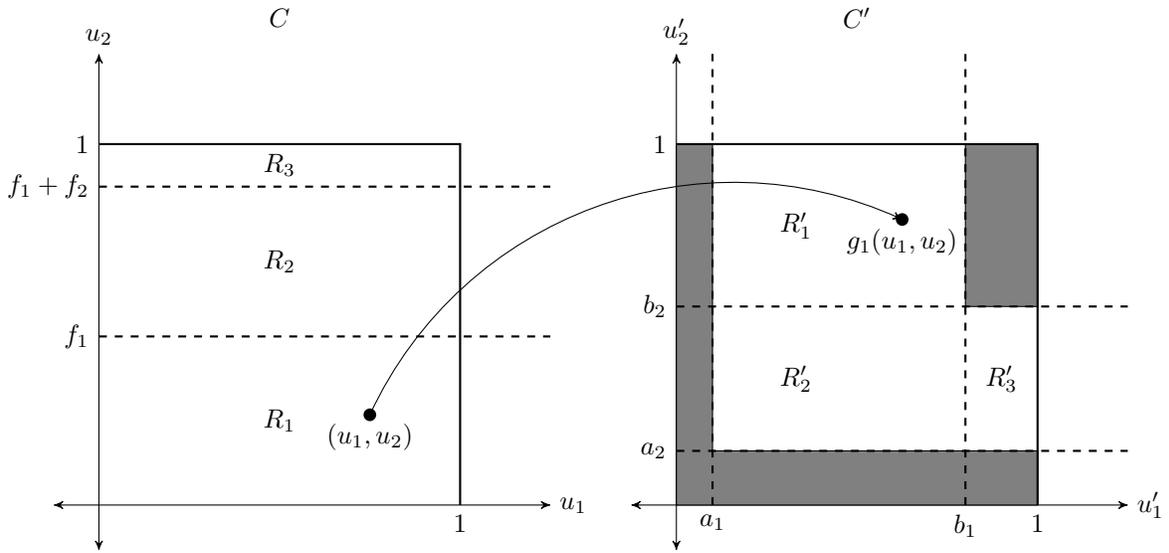}
	\caption{Unrestricted to restricted mapping for $k = 1, N = 2$.}
	\label{fig:diagram}
\end{figure}

The method is composed of the following steps. First generate a sample $(u_1, u_2)$ from the uniform distribution over the unit square. Then, based on the value of $u_2$, determine which region $R_s \subset C$ this point falls into and transform the coordinates using $u_j^{\prime} = g_s(u_j)$. These new coordinates fall within $C^{\prime}$ with the same probability distribution that rejection sampling would give but satisfies the constraints with only a single draw needed. After the draw, apply the inverse CDFs to the $(u_1^{\prime}, u_2^{\prime})$ to transform to the original two random variables with $X_j = F_j^{-1}(u_j^{\prime})$. Lastly, take the minimum to obtain a sample from our desired distribution: $Y = \min(X_1, X_2)$. Figure \ref{fig:diagram} shows a diagram of the two spaces and their divisions for some example values, graying out the regions where samples in $C^{\prime}$ cannot be located.

%$Y$ where $Y > c$ and $S(Y)$ is factorable into $N$ known survival functions then $Y = \min\{X_1, \ldots, X_N\}$ by
%$$
%\begin{aligned}
%	S_{Y \mid Y > c}(y) &= \Pr(Y > y \mid Y > c)\\ 
%			&= \Pr(\min_i(X_i) > y \mid \min_i(X_i) > c) \\ 
%			&= \Pr(X_1 > y \land X_2 > y \land \cdots \land X_N > y \mid \min_i(X_i) > c)\\ 
%			&= \Pr(X_1 > y \mid X_1 > c)\Pr(X_2 > y \mid X_2 > c) \cdots \Pr(X_N > y \mid X_N > c) \\ 
%			&= S_{X_1 \mid X_1 > c}(y)S_{X_2 \mid X_2 > c}(y)\cdots S_{X_N \mid X_N > c }(y). 
%\end{aligned} 
%$$

%In other words, if the survival function of $Y$ can be factored into a collection of independent distributions then $Y$ can be re-expressed as the minimum of that collection. Sampling from the distribution for $Y$ is complicated by the addition of the boundary condition that each $S(X_i)$ must satisfy. Often practitioners would resort to rejection sampling though this is potentially very slow if the bounded area is small (\cite{Reiher1966Jan} \cite{Rubinstein1982Jun}).

To partition the restricted space $C^{\prime}$ into regions, first note that there are three ways which the $X_i$ could satisfy the bounds: only $X_1$ is less than $B$, both $X_1$ and $X_2$ are less than $B$, or only $X_2$ is less than $B$. $B$ is the only bound mentioned because in all scenarios both random variables must be greater than $A$. Create an ordered list of the index combinations for $X_i$ which are less than $B$ in each possibility: $S = \{\{1\}, \{1, 2\}, \{2\} \}$. Assign coordinates $(u_1^{\prime}, u_2^{\prime})$ to the axes of our unit square, and divide it into three regions $R_s^{\prime}$ for each $s \in S$ where
$$
R_s^{\prime} = \left\{(u_1^{\prime}, u_2^{\prime}) : \begin{array}{cc}
	\forall j \in s, & a_j < u_j^{\prime} < b_j \\ 
	\forall j \notin s, & b_j < u_j^{\prime} < 1 
	\end{array} \right\}
$$
This defines the region $R_s^{\prime}$ to be the one where the coordinates with indices in $s \in S$ lie within their boundaries, and the other coordinates exceed them. We can see that this division of the square forms a nonoverlapping cover of $C^{\prime}$, i.e. $\bigcap \limits_{s \in S} R_s^{\prime} = \emptyset$ and $\bigcup \limits_{s \in S} R_s^{\prime} = C^{\prime}$. In order to map into these sets without disrupting the probability distribution, we need to know the relative volumes of these regions. We see that for region $R_s^{\prime}$, its volume is $V_{R_s^{\prime}} = \prod \limits_{j \in s}(b_j - a_j)\prod \limits_{k \notin s}(1 - b_k)$. Furthermore, the volume of the region we cut out of the unit square to create the region $C^{\prime}$ is $V_{\text{out}} = \prod \limits_{j = 1}^{2}(1-b_j)$, so we know that the total volume of region $C^{\prime}$ is $V_{C^{\prime}} = 1 - V_{\text{out}}$. Thus, the fraction of the total volume which is occupied by $R_s^{\prime}$ is $f_s = \frac{V_{R_s^{\prime}}}{V_{C^{\prime}}}$. Because we are working with a uniform distribution, we need this fraction to be the probability that our final uniform sample came from region $R_s^{\prime}$.

We want to start with a uniform draw over an unrestricted unit square which we will call $C$. Call the coordinates of this sample $(u_1, u_2)$, where $\forall i \in \{1, 2\}, 0 < u_i < 1$. We need to find a mapping from $C$ to $C^{\prime}$ which preserves the relative probabilities of falling into the regions $R_s^{\prime}$. To do this, we can cut along coordinate $u_2$ to create three regions $R_s$ with volumes $f_s$ in our new space so that we can map region $R_s$ into $R_s^{\prime}$ and preserve this probability. Thus, in our unrestricted space, we can define: 
$$ 
R_s = \{(u_1, u_2) : 0 < u_1 < 1; \sum \limits_{j = 1}^{s - 1}f_j < u_2 < \sum \limits_{j = 1}^{s}f_j \} 
$$
so that the total volume of $R_s$ is $f_s$. Once we have this division, it is easy to create the mapping from $R_s$ to $R_s^{\prime}$: $g_s(u_j) = u_j^{\prime}$ where

\begin{align*}
g_s(u_j) = \left\{ \begin{array}{cc} 
	(b_j - a_j)u_j + a_j & \text{if }j \in s, j = 1 \\
	\frac{b_j - a_j}{f_s}(u_j - \sum \limits_{l = 1}^{s - 1}f_l) + a_j & \text{if }j \in s, j = 2 \\ 
	(1 - b_j)u_j + b_j & \text{if }j \notin s, j = 1 \\ 
	\frac{1 - b_j}{f_s}(u_j - \sum \limits_{l = 1}^{s - 1}f_l) + b_j & \text{if }j \notin s, j = 2
	\end{array} \right.  
\end{align*}
where the function $g_s(u_j)$ indicates that this $g_s$ is only to be used to map between points in $R_s$ and $R_s^{\prime}$, where $(u_1, u_2) \in R_s$ if $\sum \limits_{j = 1}^{s - 1}f_j < u_2 < \sum \limits_{j = 1}^{s}f_j$. Define the final mapping to be $g(u_1, u_2) = (g_s(u_1), g_s(u_2))$ when $(u_1, u_2) \in R_s$.

\section{General Procedure}
The following is the general procedure for $N$ random variables of which we want the truncated $k$-th order statistic. Denote the unrestricted $N$-dimensional unit hypercube from which we will draw our original uniform sample as $C = [0,1]^N$, and let points in this space have coordinates given by $(u_1, \cdots, u_N)$. Let the bound restricted hypercube be $C^{\prime}$ with points having coordinates $(u_1^{\prime}, \cdots, u_N^{\prime})$. Our original bounds, $A$ and $B$, apply to the final $Y$ and not the uniform samples. By applying the CDFs, $a_i = F_{X_i}(A)$ and $b_i = F_{X_i}(B)$, we obtain the appropriate limits on $u_i^{\prime}$. Using these definitions, we can see that the restricted region $C^{\prime}$ from which we need to generate uniform samples can be written as:
$$
C^{\prime} = \left\{(u_1^{\prime}, \cdots, u_N^{\prime}) : \begin{array}{cc}
	 a_j < u_j^{\prime} < b_j & \text{for at least } k \text{ of the }j \in \{1, \cdots, N\} \\ 
	  0 < u_j^{\prime} < 1 & \text{otherwise}
	  \end{array} \right\} 
$$
Our procedure divides $C$ and $C^{\prime}$ into the corresponding regions, $R_s$ and $R_s^{\prime}$, and determines the mapping of points from $R_s$ to $R_s^{\prime}$. After the mapping we can use ordinary inverse transform sampling to generate our draws from $Y$.

\subsection{Space Segmentation}
Each draw must satisfy the constraint that at least $k$ of the $X_i$ must be less $B$ in order for $Y$ to be less than $B$, and at least $N-k+1$ of them must be greater than $A$ in order for $Y$ to be greater than $A$. To accomplish the task, we enumerate all the possible combinations of allowed $X_i$ orderings. To accomplish this, let us ease the notation of each $X_i$ to be represented by its index $i$, and let $I = \{1, \cdots, N\}$ as the set of all possible indices. Then, we can define two sets composed of subsets of $I$ 
\begin{align*}
  S_A &= \{ s \in \mathcal{P}(I): |s| \geq N-k+1\} \\ 
  S_B &= \{ s \in \mathcal{P}(I): |s| \geq k\}
\end{align*}
where $\mathcal{P}(I)$ is the power set of $I$. The first set, $S_A$, is the list of possible combinations of the $X_i$ which could fulfill the $A$ bound and likewise for $S_B$. However, not every pair of $s_A \in S_A$ and $s_B \in S_B$ is a valid combination. For instance, if $N = 4, k = 3$ and we have two sets $s_A = \{1, 2\}, s_B = \{1,2,4\}$, then we see that $3 \notin s_A$ and $3 \notin s_B$. This means that $X_3 < A$ and $X_3 > B$ but since $A < B$, then we have a contradiction. In order to avoid such problems, we must impose two restrictions on the pair: $s_A \cup s_B = I$ so that no index is omitted and $s_A \cap s_B \neq \emptyset$ so that there is a value satisfying the bounds to be our $k$-th order statistic. We define the set of allowed index combinations to be: 
$$ S = \{ s = (s_A, s_B): s_A \in S_A, s_B \in S_B, s_A \cup s_B = I, s_A \cap s_B \neq \emptyset \} $$ 
Assume that this set is given an order and can therefore be indexed by integers between $1$ and $|S|$.

Using the above index list, we can divide $C^{\prime}$ into smaller regions $R_s^{\prime}$ indexed by the elements $s \in S$, where 
$$
R_{s = (s_A, s_B)}^{\prime} = \left\{(u_1^{\prime}, \cdots, u_N^{\prime}) : \begin{array}{cc}
	 a_j < u_j^{\prime} < b_j & \text{if } j \in s_A \cap s_B \\ 
	 b_j < u_j^{\prime} < 1 & \text{if } j \in s_A \setminus s_B \\
	  0 < u_j^{\prime} < a_j & \text{if } j \in s_B \setminus s_A 
	  \end{array} \right\}. 
$$
Thus, $R_s^{\prime}$ is the region of the unit hypercube in which all $X_i$ for $i \in s_A$ are greater than $A$ and all $X_i$ for $i \in s_B$ are less than $B$. It is easy to see that $R_{s_1}^{\prime}$ and $R_{s_2}^{\prime}$ are non-overlapping for $s_1 \neq s_2$ since changing any index's presence in either set changes the allowed range of values. Additionally, the union of these regions is the total restricted region $C^{\prime}$ since there are no further combinations which give us a $k$-th order statistic in the allowed range of values.

In defining the regions $R_s \subset C$ corresponding to the $R_s^{\prime}$, we ensure that the probability distribution after the mapping is uniform over $C^{\prime}$ to conduct inverse transform sampling. The probability densities over both $C$ and $C^{\prime}$ follow a multivariate uniform distribution. Because the density function is inversely proportional to the volume, using our mapping we can maintain the relative volumes of the regions being mapped to one another. Then the volume of the regions in $C$, $V_{R_s}$, are equal to the normalized volumes in the restricted regions in $C^{\prime}$, $f_s$. The normalizing of the volumes begins with the un-normalized volume of $R_s^{\prime}$, $V_{R_s^{\prime}}$, which is given by 
$$
V_{R_s^{\prime}} = \prod \limits_{l \in s_A\cap s_B}(b_l - a_l) \prod \limits_{m \in s_A\setminus s_B}(1 - b_m) \prod \limits_{n \in s_B\setminus s_A}a_n 
$$ 
Then, the normalized volume is the fraction of the allowed volume occupied by the region $R_s^{\prime}$ and is given by $f_s = \frac{V_{R_s^{\prime}}}{\sum \limits_{\sigma \in S}V_{R_{\sigma}^{\prime}}}$.

Then, we can define the regions $R_s$ in $C$ so that they have volume $f_s$
$$ 
R_s = \left\{(u_1, \cdots, u_N) : \begin{array}{cc} 
	0 < u_j < 1 & \text{if }j < N \\
	\sum \limits_{l = 1}^{s - 1}f_l < u_j < \sum \limits_{l = 1}^{s}f_l & \text{if }j = N 
	\end{array} \right\}
$$
Clearly, our region $R_s$ has volume $f_s$, and, similarly to $R_s^{\prime}$, $R_{s_1}$ and $R_{s_2}$ are non-overlapping and the union of all of the $R_s$ comprises the whole space $C$. 

\subsection{Mapping Between Spaces}
Now we have defined a set of allowed orderings of the random variables as $S$ whose elements $s$ identify regions $R_s \subset C$ and $R_s^{\prime} \subset C^{\prime}$. Once we have our uniformly sampled point $(u_1, \cdots, u_N)$ from $C$, we use the above definition of the regions to determine which $R_s$ our point falls into. Once we know this $s$, we can use the following mapping to translate the point into $C^{\prime}$ coordinates $(u_1^{\prime}, \cdots, u_N^{\prime})$. For coordinate index $j$ and region index $s = (s_A, s_B)$, we have the mapping
\begin{align*} 
	u_j^{\prime} = g_s(u_j) = \left\{ \begin{array}{cc} 
	(b_j - a_j)u_j + a_j & \text{if }j \in s_A \cap s_B, j < N \\
	\frac{b_j - a_j}{f_s}(u_j - \sum \limits_{l = 1}^{s - 1}f_l) + a_j & \text{if }j \in s_A \cap s_B, j = N \\ 
	(1 - b_j)u_j + b_j & \text{if }j \in s_A \setminus s_B, j < N \\ 
	\frac{1 - b_j}{f_s}(u_j - \sum \limits_{l = 1}^{s - 1}f_l) + b_j & \text{if }j \in s_A \setminus s_B, j = N \\ 
	a_j u_j & \text{if }j \in s_B \setminus s_A, j < N \\ 
	\frac{a_j}{f_s}(u_j - \sum \limits_{l = 1}^{s - 1}f_l) & \text{if } j \in s_B \setminus s_A, j = N 
	\end{array} \right. 
\end{align*}  
After performing this mapping on each of the $N$ coordinates in our original sampled point, we obtain $(u_1^{\prime}, \cdots, u_N^{\prime})$. Using the inverse CDFs, we can use these coordinates to obtain samples from the $X_i$: $(F_{X_1}^{-1}(u_1^{\prime}), \cdots, F_{X_N}^{-1}(u_N^{\prime}))$. Then our final answer is the $k$-th largest of these values, which due to our procedure is guaranteed to be between $A$ and $B$ without biasing the distribution.

\section{Simulation}
We can demonstrate the validity and computational speed of the above method by simulating draws in R.\footnote{R code for performing these procedures is available upon request.} For the following simulations, we choose $N = 5$ and $k = 3$, and we use the following probability distributions for the individual $X_i$:
\begin{itemize}
	\item $X_1 \sim \operatorname{Cauchy}(x_0 = 5, \gamma = 1)$
	\item $X_2 \sim \operatorname{Normal}(\mu = 6, \sigma = 2)$
	\item $X_3 \sim \operatorname{Logistic}(\mu = 3, s = 2)$
	\item $X_4 \sim \operatorname{Weibull}(\lambda = 10, k = 1.5)$
	\item $X_5 \sim \operatorname{Uniform}(a = -5, b = 20)$
\end{itemize}
We perform draws using both our new method and ordinary rejection sampling. First, we can compare the empirical CDFs generated from $10000$ draws using both methods on a variety of ranges.

\begin{figure}[h!]
    \centering
%    \begin{subfigure}[b]{\textwidth}
%        \centering
%        \includegraphics[scale = 0.5]{k3N5A1B8time041719094327.pdf}
%        \caption{}
%        \label{fig:cdf18}
%    \end{subfigure}
    \begin{subfigure}[b]{\textwidth}
        \centering
        \includegraphics[scale = 0.65]{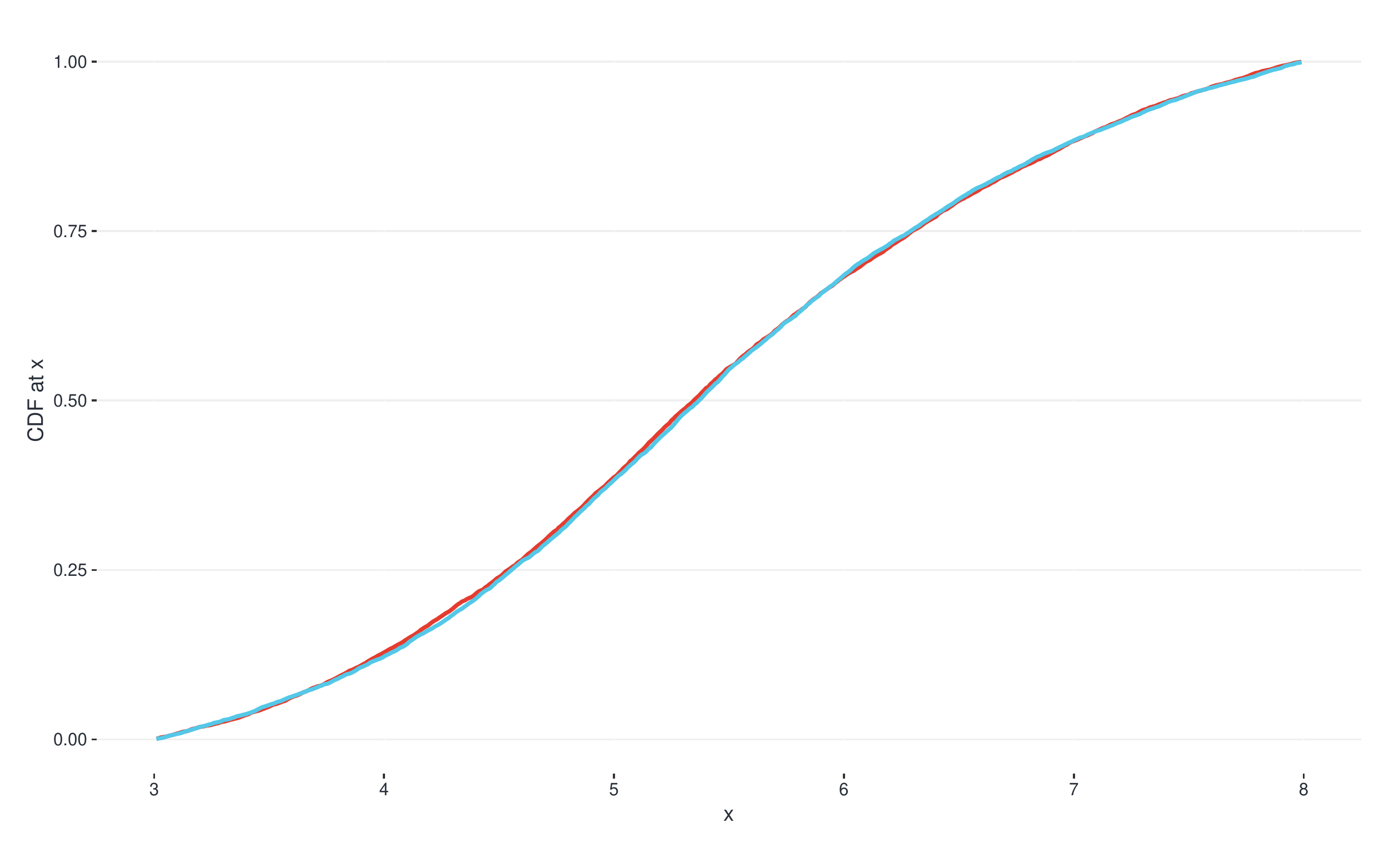}
        \caption{Empirical CDF in range $[3, 8]$}
        \label{fig:cdf38}
    \end{subfigure}
    \begin{subfigure}[b]{\textwidth}
        \centering
        \includegraphics[scale = 0.65]{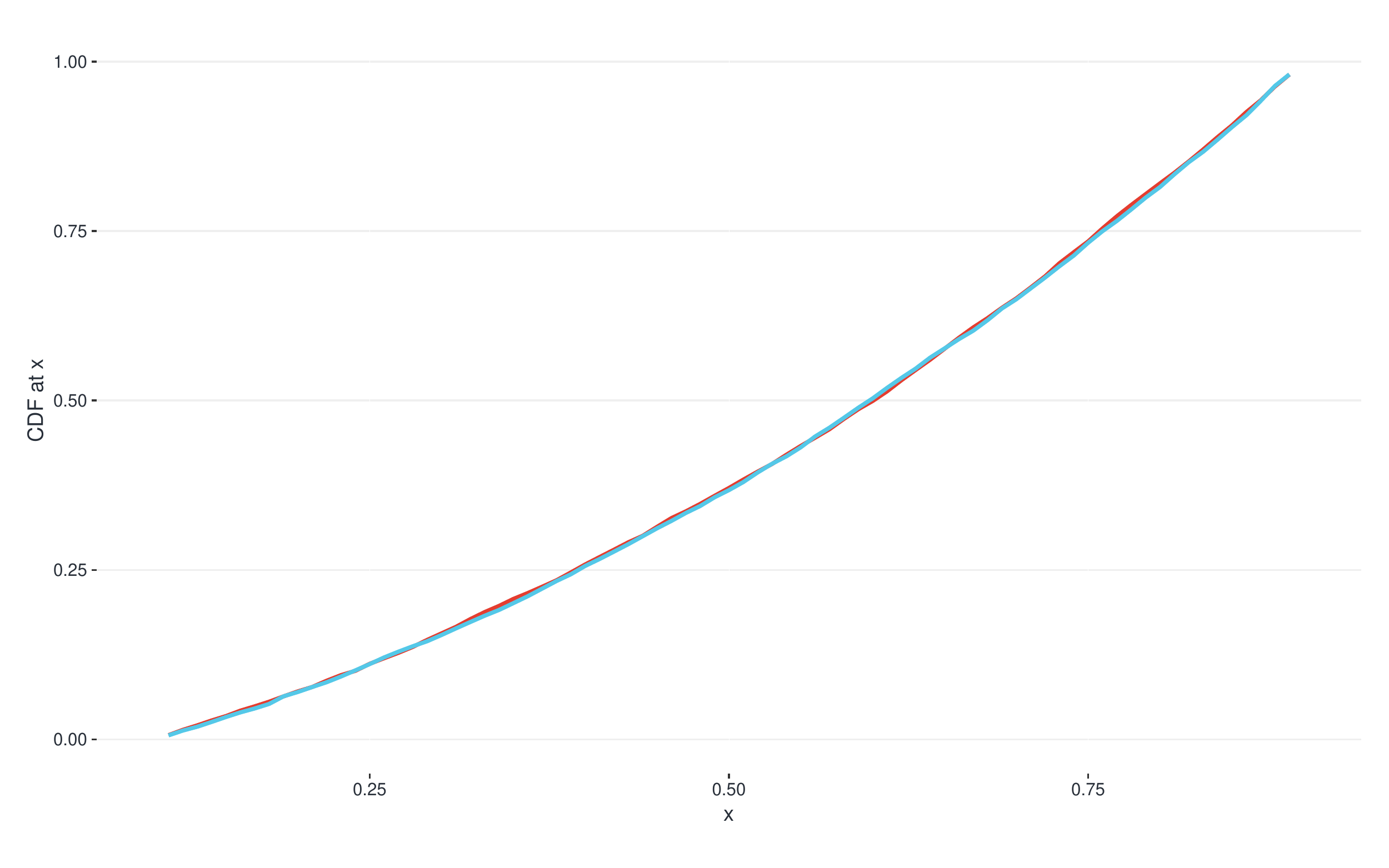}
        \caption{Empirical CDF in range $[0.1, 0.9]$}
        \label{fig:cdf01}
    \end{subfigure}
    \caption{Comparison of the empirical CDFs for three sets of bounds between our method (red) versus rejection sampling (blue)}
    \label{fig:cdfComp}
\end{figure}

In Figure \ref{fig:cdfComp}, we see that there is great agreement between the CDFs as determined by the two methods, to the level expected for $10000$ draws. Once certain that our method is accurately drawing from the truncated order statistic distribution, we can assess how long it takes relative to rejection sampling. In determining timing, we used a Dell Latitude E7450 running Windows 10 Pro with a 2.6 GHz Intel Core i7 processor and 16GB 1600 MHz DDR3 memory. For the timing comparison, we used six different intervals as bounds and sampled from each interval $1$, $10$, $100$, $1000$, and $10000$ times, comparing the time it takes to draw all of the samples.

\begin{figure}[h!]
    \centering
    \begin{subfigure}[b]{\textwidth}
        \centering
        \includegraphics[scale = 0.65]{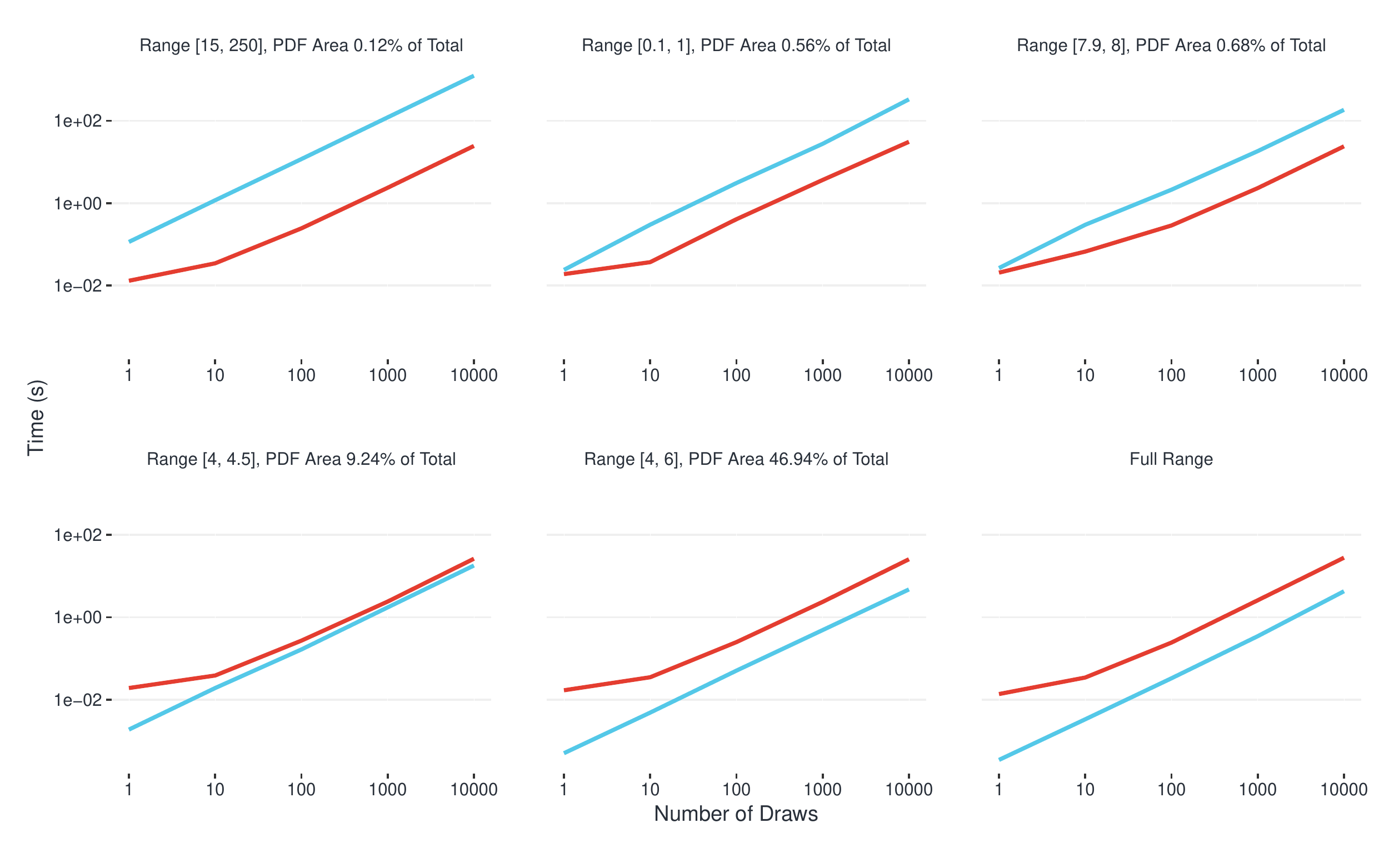}
        \caption{Time for varying numbers of draws for a selection of regions}
        \label{fig:timeIter}
    \end{subfigure}
    \begin{subfigure}[b]{\textwidth}
        \centering
        \includegraphics[scale = 0.65]{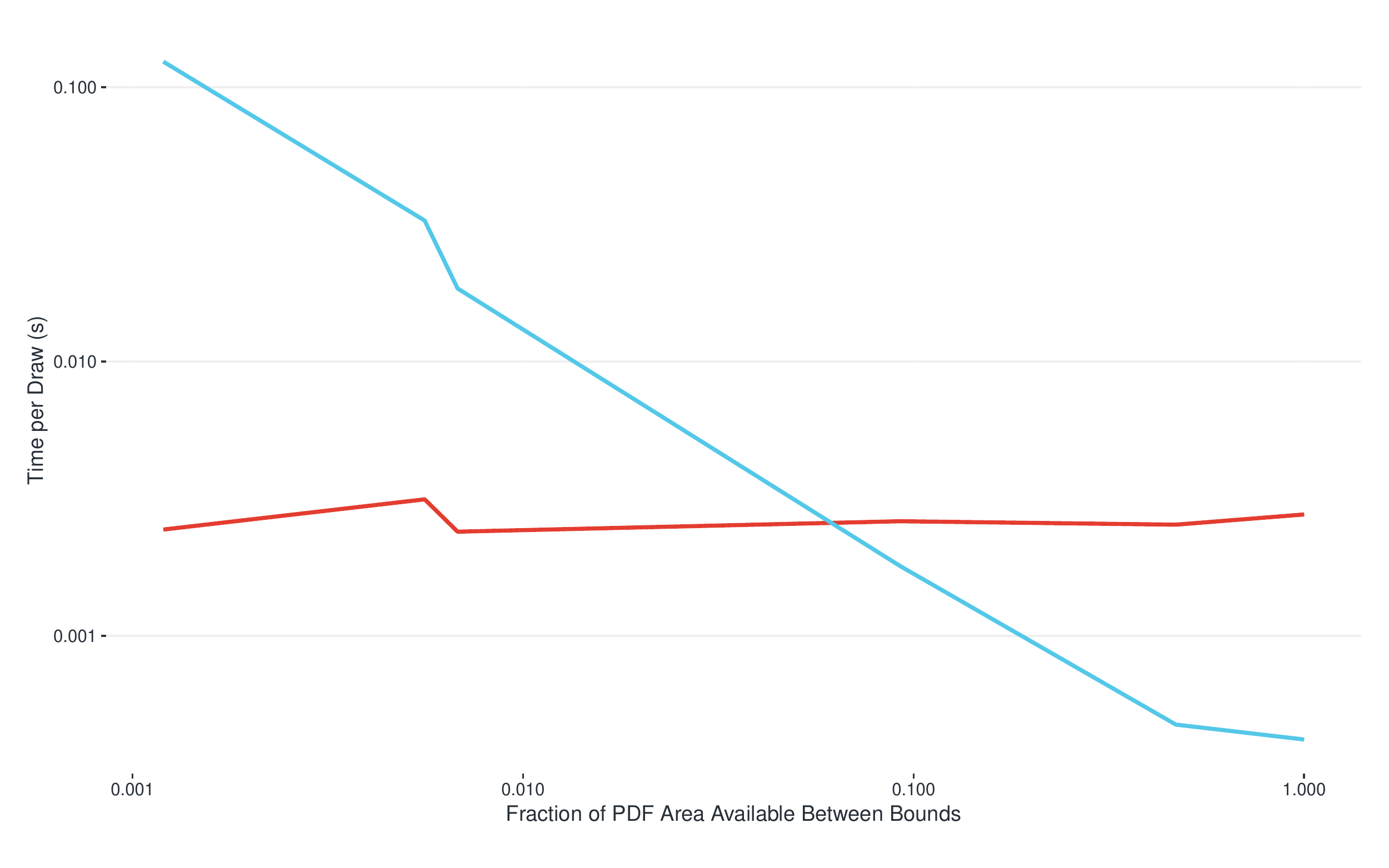}
		\caption{Average time per draw as a function of the available PDF area}        
        \label{fig:timeAvg}
    \end{subfigure}
    \caption{Comparison of times for our method (red) versus rejection sampling (blue)}
    \label{fig:timeComp}
\end{figure}

In Figure \ref{fig:timeIter}, we show that both rejection sampling and our new method behave similarly as the number of samples drawn is increased. This behavior is approximately linear, which makes sense due to the fact that any overhead time would be negligible in comparison with the amount of time per draw, though this overhead time for our new method can be seen in the relatively higher time taken for only one sample. In addition to the number of samples drawn, we need to consider the probability density between the bounds being considered.

Figure \ref{fig:timeAvg} shows how the execution time varies as the bounds are changed. The bounds themselves are not of interest here since an interval of fixed length can be very easy to sample from if near the peak of the distribution or hard to do if in the tail. Instead, the area of the PDF in that interval is what controls how long rejection sampling may take. The new method is less efficient than rejection sampling when the PDF area is above 5-10\%, but below that it can take considerably less time. The extra time taken by the new method is largely due to the extra computations involved in the mapping and determining the mapping groups. The running time of the new method should not vary with the PDF area since every draw leads to a valid sample, which is supported by the figure, but rejection sampling varies by multiple orders of magnitude as the area gets smaller due to the number of draws needed to find a valid one. If a fraction $f$ of the total PDF area is available within the bounds, then it will take $\frac{1}{f}$ draws to find a valid sample on average from rejection sampling. The smallest PDF areas used here were ~0.1\% of the total area, but as this gets lower (which could occur for narrow ranges or near the tails of a distribution) the time taken should continue to increase dramatically.

\FloatBarrier

\section{Conclusion}
By mapping unrestricted uniform draws from the $N$-dimensional unit hypercube onto a restricted subspace based on given lower and upper bounds, we can utilize inverse transform sampling to simulate draws from the $k$-th order statistic of $N$ random variables with known CDFs and inverse CDFs. This procedure results in an unbiased draw from this distribution, and in the case where the PDF density in the region of interest is low, the speed of execution is much faster than rejection sampling. This method could be used in place of rejection sampling to ensure that the execution will take a known amount of time if the density of the PDF is unknown in the bounded volume. A practical example would be in drawing from the tail of the minimum or maximum of a set of random variables which are not identically distributed; in such a case, this method would dramatically outperform rejection sampling.

\FloatBarrier
\newpage
\nocite{*}
\bibliography{bibliography} 

\end{document}